\documentclass[12pt]{article}
\usepackage{graphicx}

\newcommand \bra[1]{\left< {#1} \,\right\vert}
\newcommand \ket[1]{\left\vert\, {#1} \, \right>}

\newcommand{\bea}{\begin{eqnarray}}
\newcommand{\eea}{\end{eqnarray}}
\newcommand{\simgt}{\hbox{ \raise3pt\hbox to 0pt{$>$}\raise-3pt\hbox{$\sim$} }}
\newcommand{\simlt}{\hbox{ \raise3pt\hbox to 0pt{$<$}\raise-3pt\hbox{$\sim$} }}

\begin{document}
\begin{titlepage}
\title{Renormalon Cancellation in Heavy Quarkonia\\
and Determination of $m_b$, $m_t$%
\thanks{Based on the invited talk 
``Top quark physics at future linear colliders''
given
at the Japan Physics Society Meeting,
Osaka, Japan, March 30 - April 2, 2000.}
}
\author{Y.~Sumino
\\ \\ Department of Physics, Tohoku University\\
Sendai, 980-8578 Japan
}
\date{}
\maketitle
\thispagestyle{empty}
\vspace{-3.8truein}
\begin{flushright}
{\bf TU--588}\\
{\bf April 2000}
\end{flushright}
\vspace{3.0truein}
\vspace{3cm}
\begin{abstract}
\noindent
{\small
This is an elementary introduction to the recent
significant theoretical progress in the field
of heavy quarkonium physics.
We show how
renormalon cancellation takes place in the heavy
quarkonium system, such as bottomonium and
(remnant of) toponium resonance,
and how this notion is useful in extracting the 
$\overline{\rm MS}$ masses of the bottom and top quarks.
}
\end{abstract}
\vfil

\end{titlepage}
  
\section{Introduction}

Recently there has been significant progress in our understanding
of heavy quarkonia such as $\Upsilon$'s and remnant of toponium
resonances.
Developments in technologies of higher order calculations and the
subsequent discovery of renormalon cancellation enabled extractions
of $m_b$ and (in future experiments) of $m_t$ with high accuracy 
from the quarkonium spectra.
In this article we review the notion of renormalon and
its cancellation in the heavy quarkonium system.
We demonstrate how it is useful in extracting the quark masses.\footnote{
See Ref.~\cite{beneke} for a comprehensive review of renormalons.
}

We consider heavy quarkonia
whose sizes (given by the Bohr radius 
$\sim (\alpha_S m_q)^{-1}$) are much smaller than the
hadronization scale 
$\Lambda_{\rm QCD}^{-1} \sim (0.3~{\rm GeV})^{-1}$.
In reality the candidates are
$\Upsilon (1S)$ and (remnant of) toponium resonances,
whose sizes are 
$\sim (1.5~{\rm GeV})^{-1}$ and
$\sim (20~{\rm GeV})^{-1}$, respectively.
In such a system, gluons participating in the binding of
the $q\bar{q}$ boundstate have wavelengths much shorter than
the hadronization scale, so theoretically nature of the boundstate 
can be described well using perturbative QCD.
In particular
the boundstate spectrum (the mass of boundstate) can be
calculated as a function of the quark mass and $\alpha_S$.
Consequently we can extract the quark masses, 
$m_b$, $m_t$, from the masses of the above quarkonia.

First let us state briefly the theoretical framework used in contemporary
calculations of spectra of non-relativistic boundstates
such as the heavy quarkonia.
In old days people solved the celebrated
Bethe-Salpeter equation
to compute the boundstate spectrum.
We no longer use this equation;
instead we reduce the problem to a quantum mechanical one.
Namely we solve the non-relativistic Schr\"odinger equation
\bea
\hat{H} \, \psi_n (r) = E_n \, \psi_n (r)
\eea
to determine the boundstate wave functions and energy spectrum.
The quantum mechanical Hamiltonian is determined from
perturbative QCD order by order in expansion in $1/c$
(inverse of the speed of light):
\bea
\hat{H} =  \hat{H}_0 + \frac{1}{c} \, \hat{H}_1
+ \frac{1}{c^2} \, \hat{H}_2 + \cdots .
\label{nrhamiltonian}
\eea
Since quark and antiquark inside the heavy quarkonium are
non-relativistic, the expansion in $1/c$ leads to a
reasonable systematic approximation.
Presently the Hamiltonian is known up to ${\cal O}(1/c^2)$
\cite{py,htmy}:
\bea
\rule[-6mm]{0mm}{6mm}
\hat{H}_0 &=& \frac{\vec{p}\,^2}{m} \,
\underline{- \, C_F \,\frac{\alpha_S}{r}} , 
\label{h0}
\\ \rule[-8mm]{0mm}{6mm}
\hat{H}_1 &=& ~~~~~~
\underline{- \, C_F \,\frac{\alpha_S}{r} \cdot
\biggl( \frac{\alpha_S}{4\pi} \biggr) \cdot
\biggl\{ \beta_0 \, \log ( \mu'^2 r^2 ) } + a_1 \biggr\}
, 
\\
\hat{H}_2 &=& - \frac{\vec{p}\,^4}{4m^3} \,
\underline{- \, C_F \,\frac{\alpha_S}{r}\cdot
\biggl( \frac{\alpha_S}{4\pi} \biggr)^2 \cdot
\biggl\{ \beta_0^2 \,[ \log^2 ( \mu'^2 r^2 ) + \frac{\pi^2}{3}] 
}
+ (\beta_1+2\beta_0 a_1)\log ( \mu'^2 r^2 ) + a_2
\biggr\}
\nonumber \\ &&
+ \frac{\pi C_F \alpha_S}{m^2} \, \delta^3(\vec{r})
+\frac{3 C_F \alpha_S}{2m^2r^3} \, \vec{L}\cdot\vec{S}
- \frac{C_F \alpha_S}{2m^2r} \biggl(
\vec{p}\,^2 + \frac{1}{r^2} r_i r_j p_j p_i \biggr)
- \frac{C_A C_F \alpha_S^2}{2mr^2}
\nonumber \\ &&
- \frac{C_F \alpha_S}{2m^2}
\biggl\{ \frac{S^2}{r^3} - 3 \frac{(\vec{S}\cdot\vec{r})^2}{r^5}
- \frac{4\pi}{3}(2S^2-3) \delta^3(\vec{r}) 
\biggr\} , \label{h2}
\eea
where $m$ denotes the pole mass of the quark;
$\alpha_S \equiv \alpha_S(\mu)$;
$C_F = 4/3$, $C_A=3$ are color factors; 
$\mu' = \mu \, e^{\gamma_E}$.
The lowest-order Hamiltonian $\hat{H}_0$
is nothing but that of 
two equal-mass particles interacting via the
Coulomb potential.

In addition to the above Hamiltonians, some of the terms 
in higher order Hamiltonians $\hat{H}_n$ 
(corresponding to the underlined terms) 
are known.
From the analysis of these higher order terms, one finds that
there exists a problem in extracting the quark mass from the
boundstate spectrum.
We will first address the problem, which is known as the 
``renormalon problem'',
and then will see how it is solved.

\section{The Renormalon Problem}

The terms with underlines in Eqs.~(\ref{h0})--(\ref{h2})
stem from the running of the
coupling constant dictated by
\bea
\mu^2 \frac{d \alpha_S}{d\mu^2} = - \, \frac{\beta_0}{4\pi} \alpha_S^2 
+ \cdots ,
\eea
and all higher order terms can be determined using the 
renormalization-group equation.
Thus, in the ``large $\beta_0$ approximation'', the
potential between quark and antiquark is given by\footnote{
It is the Coulomb potential with the ``running charge'';
cf.~Eq.~(\ref{h0}).
}
\bea
V_{\beta_0}(r) = - \int
\frac{d^3 \vec{q}}{(2\pi)^3} \, e^{i \vec{q} \cdot \vec{r}}
\, C_F \, \frac{ 4 \pi \alpha_{\rm 1L}(q)}{q^2}
~~~~~~
;
~~~~~~
q \equiv |\vec{q}| ,
\eea
where the 1-loop running coupling is defined as a
perturbation series in $\alpha_S(\mu)$:
\bea
\alpha_{\rm 1L}(q) \equiv 
\alpha_S(\mu) \sum_{n=0}^\infty
\biggl\{ - \, \frac{\beta_0 \alpha_S(\mu)}{4\pi} \, \log
\biggl( \frac{q^2}{\mu^2} \biggr) \biggr\}^n .
\eea
We may not resum the geometrical series before the
Fourier integration because then the integrand exhibits
a pole at $q = \Lambda$,
\bea
\alpha_{\rm 1L}(q) ~ \to ~
\frac{\alpha_S(\mu)}{ 1 + \frac{\beta_0\alpha_S(\mu)}{4\pi} \,
\log \Bigl( \frac{q^2}{\mu^2} \Bigr)}
= \frac{4\pi/\beta_0}{ \log \Bigl( \frac{q^2}{\Lambda^2} \Bigr)}
~~
\hbox{ \hbox to -2pt{$\sim$}\raise-12pt\hbox{$\scriptstyle q\to\Lambda$} }
~~
\frac{4\pi}{\beta_0} \, \frac{\Lambda^2}{q^2-\Lambda^2} ,
\label{landaupole}
\eea
and the Fourier integral becomes ill-defined.
Here, 
\bea
\Lambda \equiv \mu \, \exp \biggl[
- \, \frac{2\pi}{\beta_0 \alpha_S(\mu)} \biggr]
\eea
is the $\mu$-independent integration constant of the
1-loop renormalization-group equation.
Therefore, the above potential $V_{\beta_0}(r)$ can only be defined as a
perturbation series in $\alpha_S(\mu)$.
(Fourier integral of each term of the series is well-defined.)

When we examine the large-order behavior of
this perturbation series,
we find that it is an asymptotic series and has an intrinsic
uncertainty 
\bea
\delta V_{\beta_0}(r) \sim \Lambda = {\cal O}(300~{\rm MeV}) .
\eea
If we want to extract the quark mass from the spectrum of boundstates,
this uncertainty in the potential is directly reflected to the
uncertainty in the quark mass.
It is because the quarkonium mass is determined as twice the quark
pole mass minus the binding energy, and an uncertainty in the
potential means an uncertainty in the binding energy.
This implies that 
we cannot determine the quark mass to an accuracy better than
${\cal O}(300~{\rm MeV})$.

Now we examine the series expansion of $V_{\beta_0}(r)$ and
see how the uncertainty arises \cite{bbb,al}.
If we perform the Fourier integration term by term,
\bea
\rule[-8mm]{0mm}{6mm}
V_{\beta_0}(r) &=& - C_F \, 4 \pi \alpha_S(\mu) \sum_{n=0}^\infty
\int
\frac{d^3 \vec{q}}{(2\pi)^3} \, \,
\frac{e^{i \vec{q} \cdot \vec{r}}}{q^2}
\, \, \biggl\{ - \, \frac{\beta_0 \alpha_S(\mu)}{4\pi} \, \log
\biggl( \frac{q^2}{\mu^2} \biggr) \biggr\}^n 
\\
&=&
- C_F \, 4 \pi \alpha_S(\mu) \sum_{n=0}^\infty \,
\biggl\{ \frac{\beta_0 \alpha_S(\mu)}{4\pi} \biggr\}^n \,
f_n(r,\mu) \times n! ,
\eea
the coefficients $f_n(r,\mu)$ can be determined from a generating 
function
\bea
\rule[-8mm]{0mm}{6mm}
F(r,\mu;u) & \equiv &
\int \frac{d^3 \vec{q}}{(2\pi)^3} \, \,
\frac{e^{i \vec{q} \cdot \vec{r}}}{q^2} \, 
\biggl( \frac{\mu^2}{q^2} \biggr)^{u}
\\ \rule[-8mm]{0mm}{6mm}
&=& \frac{(\mu r)^{2u}}{4\pi r} \,
\frac{1}{\cos (\pi u) \, \Gamma(1+2u)}
\label{analF}
\\
&=& \sum_n f_n(r,\mu) \, u^n .
\label{expF}
\eea
Using this generating function, one easily obtains the asymptotic
behavior of $f_n(r,\mu)$ for large $n$.
The large-$n$ behavior of $f_n(r,\mu)$ determines the
domain of convergence of the series expansion (\ref{expF})
at $u=0$.
\begin{figure}[tbp]
  \hspace*{\fill}
    \includegraphics[width=5cm]{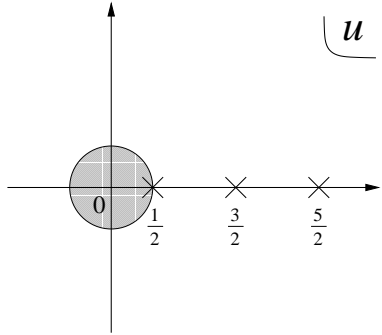}
  \hspace*{\fill}
  \\
  \hspace*{\fill}
\caption{\footnotesize
Analyticity of the generating function $F(r,\mu;u)$ 
shown on the complex $u$-plane.
Poles are located at $u=\frac{1}{2}, \frac{3}{2}, \frac{5}{2}, \cdots$.
Also the domain of convergence of the series expansion at $u=0$ is
shown.
      \label{borel}
}
  \hspace*{\fill}
\end{figure}
Conversely from the structure of the pole of (\ref{analF})
nearest to $u=0$
(see Fig.~\ref{borel}), which limits the radius of convergence,
one obtains\footnote{
The leading asymptotic behavior of $f_n(r,\mu)$ is same as
that of the expansion coefficients of
${\rm Res}[F;u=\frac{1}{2}] \times ( u - \frac{1}{2} )^{-1}$.
} 
for $n \gg 1$
\bea
f_n(r,\mu) ~\sim ~
\frac{1}{2\pi^2} \, \mu \times 2^n 
.
\eea
Note that this asymptotic behavior is independent of $r$.
This means that, although each term of 
the potential $V_{\beta_0}(r)$ is
a function of $r$, its dominant part for $n \gg 1$ is
only a constant potential which mimics the role of the quark mass
in the determination of the quarkonium spectrum.

\begin{figure}[tbp]
  \hspace*{\fill}
    \includegraphics[width=6.5cm]{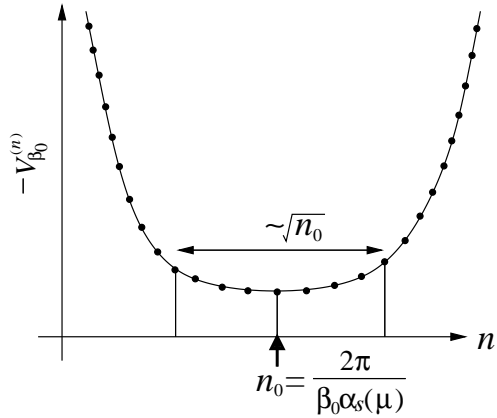}
  \hspace*{\fill}
  \\
  \hspace*{\fill}
\caption{\footnotesize
The graph showing schematically the asymptotic behavior of the $n$-th
term of $- V_{\beta_0}(r)$ for $n \gg 1$.
      \label{asympt}
}
  \hspace*{\fill}
\end{figure}

Thus, asymptotically, the $n$-th term of $V_{\beta_0}(r)$ is
given by
\bea
V^{(n)}_{\beta_0} \sim 
- C_F \, 4 \pi \alpha_S(\mu) \times\frac{\mu}{2\pi^2} \times
\biggl\{ \frac{\beta_0 \alpha_S(\mu)}{2\pi} \biggr\}^n \,
\times n! .
\eea
As we raise $n$, first the term decreases due to powers of 
the small $\alpha_S$;
for large $n$ the term increases due to the factorial $n!$.
Around $n_0 = 2\pi/(\beta_0 \alpha_S(\mu))$,
$V^{(n)}_{\beta_0}$ becomes smallest.
The size of the term scarcely changes within the range 
$n \in ( n_0 - \sqrt{n_0}, n_0 + \sqrt{n_0} )$;
see Fig.~\ref{asympt}.
We may consider the uncertainty of this asymptotic series
as the sum of the terms within this range:
\bea
\delta V_{\beta_0}(r) \sim 
\sum_{ n = n_0 - \sqrt{n_0} }^{n_0 + \sqrt{n_0}} \, 
\left| V^{(n)}_{\beta_0} \right|
\sim \Lambda .
\eea
The $\mu$-dependence vanishes in this sum, and this leads to the
claimed uncertainty.

In passing, we note that this asymptotic series is not Borel summable;
a Borel summable series has terms alternating in sign, but the 
asymptotic series originating from the QCD infrared renormalon 
has terms with the same sign.
We cannot circumvent the uncertainty by Borel summation of the series.

\section{Renormalon Cancellation in the Total Energy of a
$q\bar{q}$ system}

\begin{figure}[tbp]
  \hspace*{\fill}
    \includegraphics[width=12cm]{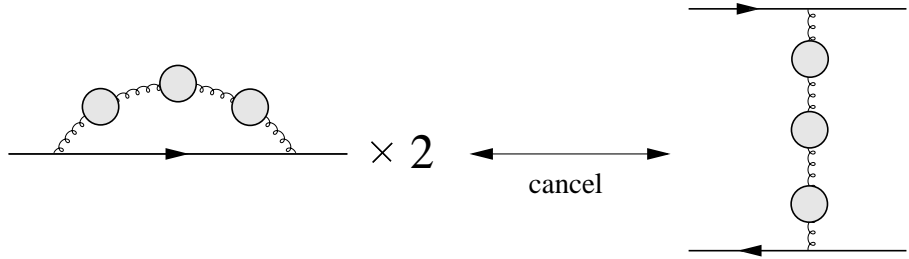}
  \hspace*{\fill}
  \\
  \hspace*{\fill}
\caption{\footnotesize
The 1-loop quark self-energy diagram and the one-gluon-exchange
diagram between quark and antiquark, where the gluon propagators
are replaced by the bubble-chains.
These diagrams, respectively, contribute to the renormalons
in $2 m_{\rm pole}$ and $V_{\beta_0}(r)$.
      \label{renormcancel}
}
  \hspace*{\fill}
\end{figure}

Now we state how the problem can be circumvented.
Consider the total energy of a color-singlet
non-relativistic quark-antiquark pair:
\bea
E_{\rm tot}(r) \simeq 2 m_{\rm pole} + V_{\beta_0}(r) .
\eea
It was found \cite{renormalon} 
that the leading renormalon contained in the potential $V_{\beta_0}(r)$ 
gets cancelled in the total energy $E_{\rm tot}(r)$
if the pole mass $m_{\rm pole}$ is expressed in terms of
the $\overline{\rm MS}$ mass.
The potential and the pole mass are expressed in terms of 
the 1-loop running coupling $\alpha_{\rm 1L}(q)$ as
\begin{eqnarray}
&&
V_{\beta_0}(r) = - \int 
\frac{d^3\vec{q}}{(2\pi)^3} 
\, e^{i \vec{q} \cdot \vec{r}} \, 
C_F \frac{4\pi\alpha_{\rm 1L}(q)}{q^2} ,
\\
&&
m_{\rm pole} \simeq m_{\overline {\rm MS}}(\mu ) +
\frac{1}{2} {\hbox to 18pt{
\hbox to -5pt{$\displaystyle \int$} 
\raise-15pt\hbox{$\scriptstyle {q}< \mu$} 
}}
\frac{d^3\vec{q}}{(2\pi)^3} \, 
C_F \frac{4\pi\alpha_{\rm 1L}(q)}{q^2} .
\end{eqnarray}
The potential $V_{\beta_0}(r)$ is essentially the Fourier transform of the
Coulomb gluon propagator exchanged between quark and antiquark;
the difference of $m_{\rm pole}$ and $m_{\overline {\rm MS}}$ is 
essentially the infrared portion of the quark 
self-energy, see Fig.~\ref{renormcancel}.
As we saw, the renormalon uncertainty is related to the ``would-be
pole'' contained in $\alpha_{\rm 1L}(q)$, cf.\ Eq.~(\ref{landaupole}).
The signs of the renormalon contributions are opposite
between $V_{\beta_0}(r)$ and $m_{\rm pole}$ because the color charges are
opposite between quark and antiquark while the self-enregy is
proportional to the square of a same charge.
Their magnitudes differ by a factor of two because
both the quark and antiquark propagator poles contribute in the
calculation of the potential whereas only one of the two contributes
in the calculation of the self-energy.
Expanding the Fourier factor in $2m_{\rm pole}$
in a Taylor series for small ${q}$, 
\bea
e^{i \vec{q} \cdot \vec{r}} = 1 + {i \vec{q} \cdot \vec{r}}
+ \frac{1}{2}({i \vec{q} \cdot \vec{r}})^2 + \cdots ,
\eea
the would-be pole contained in the
leading term gets cancelled against $V_{\beta_0}(r)$\footnote{
We are interested in the infrared region $q \sim \Lambda$.
The expansion is justified since
the typcal distance between quark and antiquark is much smaller
than the hadronization scale 
$\left< \Lambda {r} \right> \ll 1$. 
}
and consequently the
renormalon contributions cancel.

As a result of this cancellation, 
the series expansion of the total energy in $\alpha_S(\mu)$ behaves
better if we use the $\overline{\rm MS}$ mass instead of
the pole mass.
Residual uncertainty due to uncancelled pole can be estimated
similarly as in the previous section and is suppressed as
\bea
\Lambda \times \left< ({\vec{q} \cdot \vec{r}})^2 \right>
\sim \Lambda \times
\biggl( \frac{\Lambda}{\alpha_S m_{\rm pole}} \biggr)^2 ,
\eea
which is much smaller than the original uncertainty.

\section{
Extracting the $\overline{\rm MS}$ masses using the 
Full NNLO Result
}

To see how well the renormalon cancellation works, we examine
extractions of the bottom and top quark masses using the
full next-to-next-to-leading order (NNLO) result of the
boundstate spectrum.
The full NNLO formula for the lowest lying ($1S$) boundstate can be
calculated from the Hamiltonian Eqs.~(\ref{h0})-(\ref{h2}) and is
given by \cite{py,my}
\bea
M_{1S} = 2 m_{\rm pole} - \frac{4}{9} \alpha_S(\mu)^2 m_{\rm pole}
\Biggl[ 
{\textstyle
\, 1 + \frac{\alpha_S(\mu)}{\pi}
\biggl\{ \Bigl( {\scriptstyle 11} - \frac{2}{3} n_l \Bigr) L + 
\Bigl( \frac{97}{6}-\frac{11}{9}n_l \Bigr) \biggr\}
}
~~~~~~~~~~~~~
\nonumber \\
{\textstyle
+ \Bigl( \frac{\alpha_S(\mu)}{\pi} \Bigr)^2
\biggl\{ \Bigl( \frac{363}{4} - {\scriptstyle 11} n_l + \frac{1}{3}n_l^2 \Bigr) L^2
+ \Bigl( \frac{927}{4}-\frac{193}{6}n_l+n_l^2 \Bigr) L
}
~~~~~
\nonumber \\
{\textstyle
+ \Bigl( \frac{1793}{12} + \frac{2917\pi^2}{216}
-\frac{9\pi^4}{32}+\frac{275 \zeta_3}{4} \Bigr)
+ \Bigl( - \frac{1693}{72}-\frac{11\pi^2}{18}-\frac{19\zeta_3}{2} 
\Bigr)n_l
+ \Bigl( \frac{77}{108}+\frac{\pi^2}{54}+\frac{2\zeta_3}{9} \Bigr)
n_l^2
\biggr\} \Biggr]
} ,
\label{nnlo}
\eea
where $L \equiv \log [\mu/(C_F\alpha_S(\mu) m_{\rm pole} )]$, and
$n_l$ denotes the number of massless quarks.
We may examine the size of each term of the above perturbation series.
Alternatively we may rewrite the above expression in terms of 
the $\overline{\rm MS}$ mass and examine the series.
Presently the relation between $m_{\rm pole}$ and 
$\overline{m} \equiv m_{\overline{\rm MS}}(m_{\overline{\rm MS}})$
is known up to three-loop order \cite{polemass}:
\bea
m_{\rm pole} = \overline{m}\times
\biggl[ \, 1 + \frac{4}{3} \Bigl(
{\textstyle \frac{\alpha_S(\overline{m})}{\pi}} \Bigr)
+ \Bigl(
{\textstyle \frac{\alpha_S(\overline{m})}{\pi}} \Bigr)^2
(-1.0414 n_l + 13.4434) 
~~~~~~
\nonumber \\ 
+ \Bigl(
{\textstyle \frac{\alpha_S(\overline{m})}{\pi}} \Bigr)^3
(0.6527 n_l^2 - 26.655 n_l + 190.595)
\biggr] .
\label{polemsbar}
\eea

First we apply the formula
to the $\Upsilon(1S)$ state, for which $n_l=4$.
Taking the input parameter as $m_{\rm pole} = 4.96$~GeV/%
$\overline{m} = 4.22$~GeV and setting $\mu = \overline{m}$
(i.e.\ expansion parameter is $\alpha_S(\overline{m})=0.22$),
\bea
M_{\Upsilon(1S)} &=& 
2 \times ( 4.96 - 0.05 - 0.08 - 0.11 )~{\rm GeV}
~~~~~
\mbox{(Pole-mass scheme)}
\\
&=&
2 \times ( 4.22 + 0.35 + 0.12 + 0.04 )~{\rm GeV}
~~~~~
\mbox{($\overline{\rm MS}$-scheme)} .
\label{upsilonmsbar}
\eea
One sees that the series is not at all converging in the
pole-mass scheme, whereas in the $\overline{\rm MS}$-scheme
the series is converging quite nicely up to the calculated order.
(See Sec.~\ref{s6} for details of how we derived the series in
the $\overline{\rm MS}$-scheme.)
Comparing this with the experimental value 
$M_{\Upsilon(1S)} = 9.46037 \pm 0.00021$~GeV, 
one may extract the $\overline{\rm MS}$
bottom quark mass
\bea
\overline{m}_b \equiv  m_{\overline{\rm MS}}(m_{\overline{\rm MS}})
= 4.22 \pm 0.08~{\rm GeV} .
\eea
One might think that, looking
at the behavior of the above series, we may assign a smaller 
theoretical uncertainty. 
The present uncertainty is, however,
dominated by non-perturbative uncertainties other than the
renormalon contributions.
Thus, presently $\overline{m}_b$ is determined to 2\% accuracy
\cite{upsilonmass,pro}.
It seems to be fairly good in view of the fact that its major part
is controlled by perturbative QCD.

\begin{figure}[tbp]
  \hspace*{\fill}
  \begin{minipage}{5.0cm}\centering
    \hspace*{-2.3cm}
    \includegraphics[width=8cm]{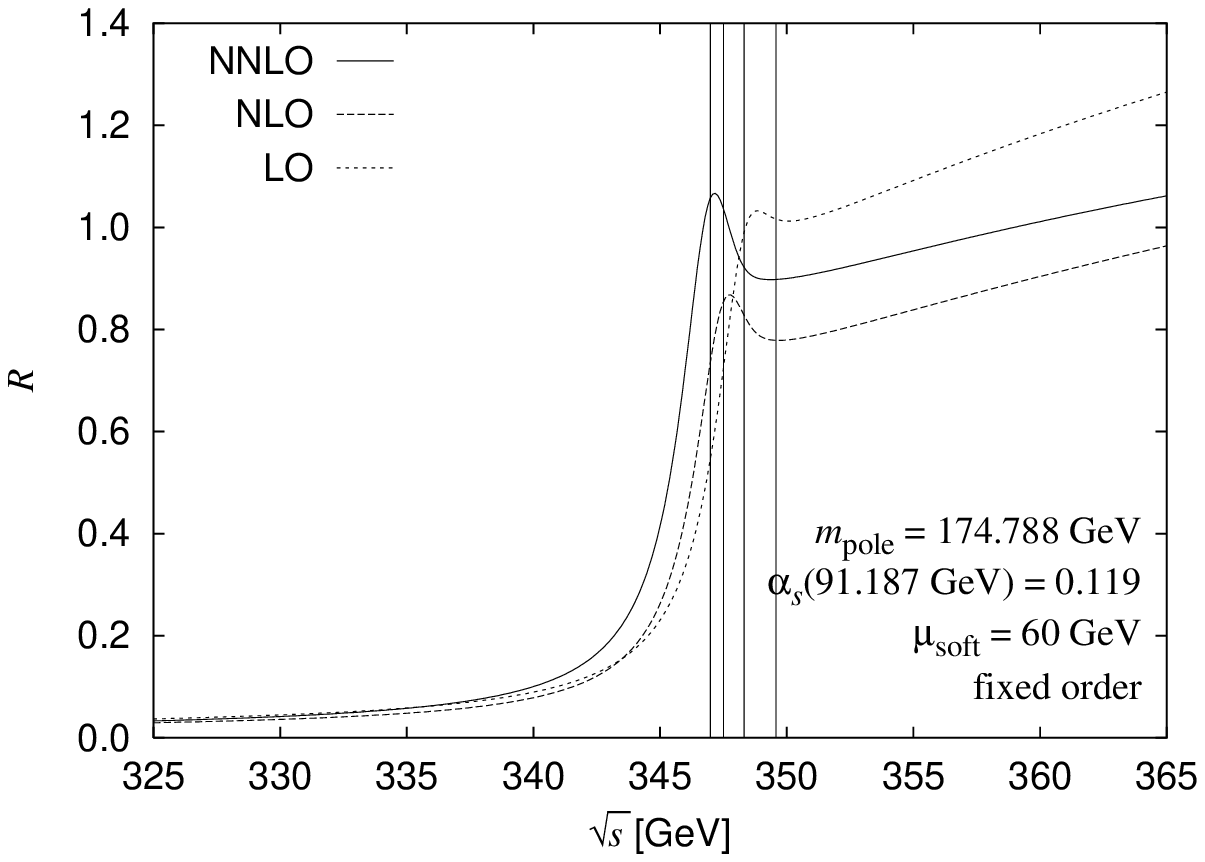}
    \hspace*{-1cm}(a)
  \end{minipage}
  \hspace*{\fill}
  \begin{minipage}{5.0cm}\centering
    \hspace*{-1cm}
    \includegraphics[width=8cm]{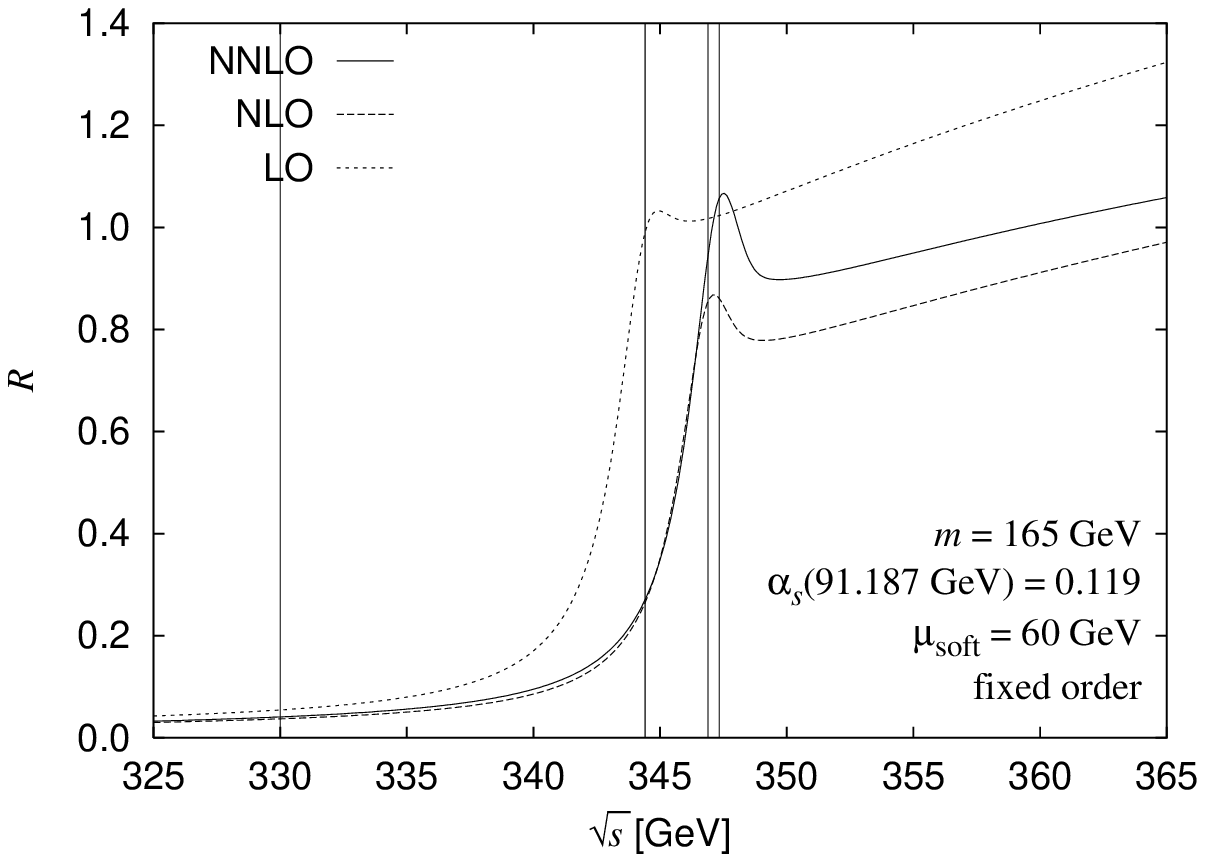}
    \hspace*{1.2cm}(b)
  \end{minipage}
  \hspace*{\fill}
  \\
  \hspace*{\fill}
\caption{\footnotesize
The total production cross sections for $e^+e^- \to t\bar{t}$
in the threshold region, where the leading-order (LO),
next-to-leading-order (NLO) and next-to-next-to-leading-order (NNLO)
curves are shown.
The vertical lines represent the locations of the corresponding
$1S$ resonances when the top width is artificially taken to zero;
also the position of $2 m_t$ is shown by vertical lines.
The two figures correspond to (a) the pole-mass scheme, and 
(b) the $\overline{\rm MS}$-scheme.
These figures are made by T.~Nagano.
      \label{nagano}
}
  \hspace*{\fill}
\end{figure}

Next we turn to the (remnant of) ``toponium'', for which
$n_l=5$.
At future linear $e^+e^-$ or $\mu^+\mu^-$ colliders, the
top quark mass will be determined to high accuracy from
the shape of the $t\bar{t}$ total production cross section 
in the threshold region.
The location of a sharp rise of the cross section is determined
mainly from the mass of the lowest lying ($1S$) $t\bar{t}$ resonance,
so we will be able to measure the resonance mass and extract
the top quark mass.
Similarly to the previous case, we set $m_{\rm pole} = 174.79$~GeV/%
$\overline{m} = 165.00$~GeV, $\alpha_S(\overline{m})=0.1091$ and obtain
\bea
M_{1S} &=& 
2 \times ( 174.79 - 0.46 - 0.40 - 0.28 )~{\rm GeV}
~~~~~
\mbox{(Pole-mass scheme)}
\\
&=&
2 \times ( 165.00 + 7.20 + 1.24 + 0.22 )~{\rm GeV}
~~~~~
\mbox{($\overline{\rm MS}$-scheme)} .
\label{mtmsbar}
\eea
In Figs.~\ref{nagano}
are shown the convergence properties of the above series
together with the corresponding cross sections.
In the pole-mass scheme, the convergence is very slow.
According to the renormalon argument, also 
uncalculated higher order terms 
would not become much smaller.
On the other hand, in the $\overline{\rm MS}$-scheme the series
shows a healthy convergence behavior.
For top quark, non-perturbative uncertainties are much
smaller than the present perturbative theoretical uncertainty.
Thus, from the above series we estimate 
that $\overline{m}_t$ can be
determined to around 100~MeV accuracy \cite{topcollab}.

\section{Physical Implications}

Let us discuss some physical implications of 
renormalon cancellation in the heavy quarkonium system.
Firstly, as already mentioned, we expect that gluons with wavelength
much longer than the size of the quarkonium cannot couple
to this system.
Hence, we expect that infrared gluons with momentum transfer
$q \ll \alpha_S m$ should decouple from the expression of
$E_{\rm tot}$.
This is a naive expectation based on classical dynamics.
Such understanding should be valid when it is described by
the bare QCD Lagrangian without large quantum corrections.
The $\overline{\rm MS}$ mass is closely related to the
bare mass of quark; only ultraviolet divergences are
subtracted.
On the other hand, the pole mass has much more intricate relation
to the bare mass, because the relation includes in addition infrared
dynamics of the quantum correction to the quark self-energy.
In this sense it would be natural to expect the decoupling
phenomenon to be realized when the 
$\overline{\rm MS}$ mass is used to express $E_{\rm tot}$.

Secondly, the pole mass of a quark is ill-defined beyond
perturbation theory.
It can be determined only when the quark can propagate 
an infinite distance.
Generally accepted belief is that when quark and antiquark are separated
beyond a distance $\sim \Lambda^{-1}$ the color flux 
is spanned between the two charges due to non-perturbative effects 
and the free quark picture is no longer valid.
On the other hand, it is natural to consider the total
energy (or the mass) of a quarkonium which is
a color-singlet state.
It can propagate for a long time and the notion of its
mass is not limited by the hadronization scale.

Thirdly, renormalon cancellation seems to be a universal feature
which occurs process independently.
The same phenomenon was known e.g.\ in the QCD corrections
to the $\rho$-parameter \cite{rho} and $B$ decays \cite{bdecay}.
Therefore, 
the $\overline{\rm MS}$ mass, which is determined accurately
from the quarkonium spectum,
would be more suited than the pole mass for an input parameter in
describing other physical processes.

\section{How to Cancel Renormalons in the Quarkonium Spectrum}
\label{s6}

\begin{figure}[tbp]
  \hspace*{\fill}
    \includegraphics[width=17cm]{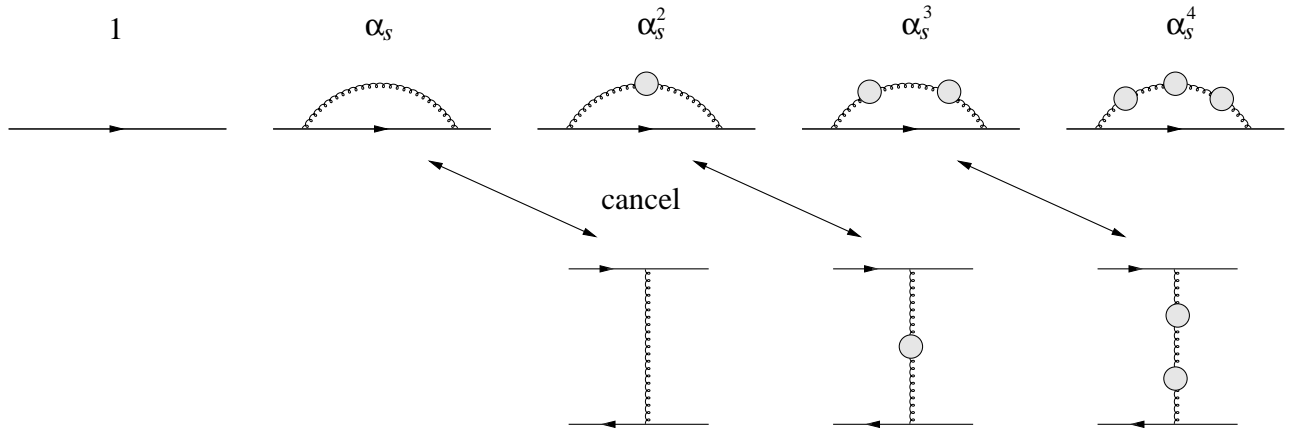}
  \hspace*{\fill}
  \\
  \hspace*{\fill}
\caption{\footnotesize
The figure showing how cancellations should take place between the
diagrams.
The orders of the potential graphs are shifted by one power of
$\alpha_S$ which is provided by the inverse of the Bohr radius
$\left< \frac{1}{r} \right> \sim \alpha_S m$.
      \label{shiftorder}
}
  \hspace*{\fill}
\end{figure}

There is one non-trivial point in realizing renormalon cancellation
in the perturbation series of the quarkonium spectrum.
When the pole mass and the binding energy are given as 
series in $\alpha_S$, renormalon cancellation takes place between
the terms whose orders in $\alpha_S$ differ by one \cite{hlm}:
\bea
\begin{minipage}[c]{15cm}
\vspace{2mm}
  \begin{flushleft}
\begin{picture}(100,60)(0,0)
\put(60,42){$2 m_{\rm pole} = 2 \overline{m} \, \,
( 1 + A_1 \, \alpha_S + A_2 \, \alpha_S^2 + A_3 \, \alpha_S^3 
+ A_4 \, \alpha_S^4 + \cdots ),$}
\put(60,0){$E_{\rm bin} ~~~ = 2 \overline{m} \, \,
( ~~~~~~~~~~~~~~~~
B_2 \, \alpha_S^2 + B_3 \, \alpha_S^3 
+ B_4 \, \alpha_S^4 + \cdots ),$}
\put(235,23){\footnotesize cancel}
\put(190,24){\vector(-3,2){20}}
\put(190,24){\vector(3,-2){20}}
\put(275,24){\vector(-3,2){20}}
\put(275,24){\vector(3,-2){20}}
\put(230,24){\vector(-3,2){20}}
\put(230,24){\vector(3,-2){20}}
\put(320,24){\vector(-3,2){20}}
\put(320,24){\vector(3,-2){20}}
\end{picture}
\end{flushleft}
\vspace{2mm}
\end{minipage}
\label{cancel}
\eea
cf.~Eqs.~(\ref{nnlo}) and (\ref{polemsbar}).
Intuitively this can be seen from the diagrams shown in Fig.~\ref{shiftorder}.
An additional power of $\alpha_S$ in the binding energy is provided
by the inverse of the Bohr radius 
$\left< r^{-1} \right> \sim \alpha_S m$,
\bea
\mbox{i.e.}~~~~~~~~~~~~~~
\left< C_F\frac{\alpha_S^n}{r} \right> \sim \alpha_S^{n+1} m .
\eea

Still, one might wonder how cancellation can ever take place between
different orders in $\alpha_S$ for any value of $\alpha_S$.
So we demonstrate the cancellations at large orders in a specific
example.
The mass of the $1S$ boundstate can be written in the form of
an expectation value as
\bea
M_{1S} = \bra{1S} 2 m_{\rm pole} + \hat{H} \ket{1S}
\eea
using the Hamiltonian (\ref{nrhamiltonian}) and 
its 1S energy eigenstate $\ket{1S}$.
We are interested in the leading renormalon contributions, so
we replace 
\bea2 m_{\rm pole} + \hat{H}
~~~\to~~~
2 m_{\rm pole} + V_{\beta_0}(r) .
\eea
The energy eigenstate can be expanded in $1/c$:
\bea
\ket{1S} = |\, {1S}^{(0)}\rangle + \frac{1}{c} |\, {1S}^{(1)}\rangle
+ \frac{1}{c^2} |\, {1S}^{(2)}\rangle + \cdots
\eea
Renormalon cancellation takes place in arbitrary
combination of $\langle{1S}^{(i)}| \cdots |{1S}^{(j)}\rangle$,
but for simplicity we evaluate only the following part:
\bea
M_{1S}^{(0)} =
\langle{1S}^{(0)}| \, 2 m_{\rm pole} + V_{\beta_0}(r) \, 
|{1S}^{(0)}\rangle
= 2 m_{\rm pole} + 
\langle{1S}^{(0)}| \, V_{\beta_0}(r) \, |{1S}^{(0)}\rangle .
\label{expec}
\eea
The second term corresponds to the binding energy and
we may evaluate it at each order of the
perturbation series:
\bea
\langle{1S}^{(0)}| V_{\beta_0}(r) |{1S}^{(0)}\rangle
&=&
\int_0^\infty dr \, r^2 \, |R(r)|^2 \, V_{\beta_0}(r) 
\nonumber \\
&=&
- \frac{1}{2} \, C_F^2 \, \alpha_S(\mu)^2 \, m_{\rm pole}
\times 
\sum_{n=0}^\infty \,
\biggl\{ \frac{\beta_0 \alpha_S(\mu)}{4\pi} \biggr\}^n \,
g_n(\mu a_0) \times n! ,
\eea
where the zeroth-order $1S$ Coulomb wave function 
is given by
\bea
R(r) = \frac{2}{a_0^{3/2}} \, e^{-r/a_0}
~~,~~~~~~~~
a_0 = 
\biggl( {\textstyle \frac{1}{2}} C_F \alpha_S(\mu) m_{\rm pole} 
\biggr)^{-1}
~:~ \mbox{Bohr radius} .
\eea
$g_n(\mu a_0)$'s are polynomials of $\log(\mu a_0)$.
Using the generating function method, one obtains the asymptotic form
\bea
g_n(\mu a_0) \sim \frac{2}{\pi} \, \mu a_0 \times 2^n 
\propto \frac{1}{\alpha_S} .
\eea
Thus, for $n \gg 1$, it becomes proportional to $\alpha_S^{-1}$
and effectively shifts the order of $\alpha_S$.
By setting $\mu = \overline{m}$, it is easy to check 
that in this example the leading renormalon cancels 
as in Eq.~(\ref{cancel}) and
the residual piece behaves as 
\bea
\Bigl[
\mbox{$n$-th term of}~M_{1S}^{(0)} \Bigr]
\sim
\alpha_S(\overline{m})\,  \overline{m} \times n! \times 
\biggl\{ \frac{\beta_0 \alpha_S(\overline{m})}{6\pi} \biggr\}^n .
\eea
It follows that
\bea
\delta M_{1S}^{(0)} \sim \Lambda \times 
\biggl( \frac{\Lambda}{\alpha_S \overline{m}} \biggr)^2 
\ll \Lambda .
\eea

From this example one learns that the cancellation 
of renormalon contribution
between shifted orders $A_n \, \alpha_S^n$ and 
$B_{n+1}\, \alpha_S^{n+1}$ should be
properly taken into account when expressing the boundstate
mass as a perturbation series.
There are many different prescriptions to accomplish
this.
We derived the
series (\ref{upsilonmsbar}) and (\ref{mtmsbar}) in the following manner.
We have rewritten Eq.~(\ref{nnlo}) as
\bea
M_{1S} &=& 2 m_{\rm pole} \times
\biggl[ \, 1 + \sum_{n=2}^4 \, P_n \, \alpha_S(\overline{m})^n
\biggr]
\nonumber \\
&=&
2 \overline{m} \times 
\biggl[ \, 1 + \sum_{n=1}^3 \, A_n \, \alpha_S(\overline{m})^n
\biggr] \times
\biggl[ \, 1 + \sum_{n=2}^4 \, P_n \, \alpha_S(\overline{m})^n
\biggr] ,
\eea
where $P_n$'s are polynomials of 
$\log \Bigl( \alpha_S(\overline{m})\Bigr)$ and
$A_n$'s are just constants independent of $\alpha_S(\overline{m})$.
We identified
$P_n \alpha_S^n$ as order $\alpha_S^{n-1}$
and then reduced the last line to a single series in $\alpha_S$.

Parametric accuracy of the last terms of 
Eqs.~(\ref{upsilonmsbar}) and (\ref{mtmsbar})
is $\alpha_S^3 \overline{m}$.
In order to improve the accuracy to $\alpha_S^4 \overline{m}$,
we need to know further 
(a) the exact 4-loop relation between $m_{\rm pole}$ and $\overline{m}$,
and 
(b) the binding energy at order $\alpha_S^5$ in the large-$\beta_0$
approximation.

\section{Some Questions}

One can ask some interesting questions related to the renormalon
problem and extraction of quark mass.
In the case of top quark, its mass will also be measured
from the invariant mass distribution of decay products
of the top quark in future hadron collider experiments and in 
future $e^+e^-$ collider
experiments.
What is the mass extracted from the peak position of the Breit-Wigner
distribution?
Naively it is the pole mass because
the peak position is determined from the position of the pole
of the top quark propagator in perturbative QCD.
On the other hand, as we have seen, the pole mass suffers from a
theoretical uncertainty of ${\cal O}$(300~MeV).
Since future experiments will be able to determine the peak position
of the invariant mass distribution to ${\cal O}$(100~MeV) accuracy, we 
will indeed face this serious conceptual problem.
We believe that renormalon cancellation also takes place in this physical
quantity.
But the problem lies in the fact that we have no reliable theoretical method
to calculate the invariant mass distribution of 
realistic color-singlet final states.
Rather we calculate the invariant mass of final partons which do not
combine to color-singlet state.
Another question is whether the peak position of
invariant mass distribution measured in hadron collider experiments
will be the same as that measured in $e^+e^-$ collider experiments.
The answer is probably no, since at hadron colliders top and
antitop are pair-created not necessarily in a color-singlet state.

These are very interesting questions which are worth further studies.

\section*{Acknowledgements}
The author is grateful to K.~Melnikov, A.~Hoang, M.~Tanabashi and
H.~Ishikawa for fruitful discussions.
Part of this work is based on a collaboration with T.~Nagano.

\def\app#1#2#3{{\it Acta~Phys.~Polonica~}{\bf B #1} (#2) #3}
\def\apa#1#2#3{{\it Acta Physica Austriaca~}{\bf#1} (#2) #3}
\def\npb#1#2#3{{\it Nucl.~Phys.~}{\bf B #1} (#2) #3}
\def\plb#1#2#3{{\it Phys.~Lett.~}{\bf B #1} (#2) #3}
\def\prd#1#2#3{{\it Phys.~Rev.~}{\bf D #1} (#2) #3}
\def\pR#1#2#3{{\it Phys.~Rev.~}{\bf #1} (#2) #3}
\def\prl#1#2#3{{\it Phys.~Rev.~Lett.~}{\bf #1} (#2) #3}
\def\sovnp#1#2#3{{\it Sov.~J.~Nucl.~Phys.~}{\bf #1} (#2) #3}
\def\yadfiz#1#2#3{{\it Yad.~Fiz.~}{\bf #1} (#2) #3}
\def\jetp#1#2#3{{\it JETP~Lett.~}{\bf #1} (#2) #3}
\def\zpc#1#2#3{{\it Z.~Phys.~}{\bf C #1} (#2) #3}

\end{document}